%% file: manuscript.tex
\documentclass[10pt,twocolumn]{article}

\usepackage{titlesec}
\usepackage[format=plain,labelfont=it]{caption}
\usepackage[left=1.5cm,right=1.5cm,top=2cm,bottom=2cm]{geometry}
\usepackage{xcolor}

\titleformat{\section}{\centering\normalfont\scshape}{\arabic{section}.}{5pt}{}
\titleformat{\subsection}{\normalfont\it}{\arabic{section}.\arabic{subsection}}{5pt}{}
\titleformat{\subsubsection}{\normalfont\it}{\arabic{section}.\arabic{subsection}.\arabic{subsubsection}}{5pt}{}

\usepackage{graphicx}
\usepackage{booktabs}
\usepackage{natbib}         %
\usepackage{amsmath,amssymb}
\usepackage{amsthm}
\newtheorem{assum}{Assumption}
\newtheorem{rem}{Remark}

\usepackage{stmaryrd}       %
\usepackage{enumitem}

\usepackage[hidelinks]{hyperref}  %
\usepackage{cleveref}

\newcommand{\R}{\mathbb{R}}

\newcommand{\N}{\mathbb{N}}

\newcommand{\Ib}{\boldsymbol{I}}

\newcommand{\Cc}{\mathcal{C}}

\newcommand{\Dk}{\Delta k}
\newcommand{\Dx}{\Delta x}

\newcommand{\ctSpace}{\mathfrak{C}}
\newcommand{\ptSpace}{\mathfrak{M}}
\newcommand{\cipher}[1]{\llbracket{}#1\rrbracket}
\newcommand{\batchsize}{\mathfrak{n}}

\newcommand{\sk}{\mathtt{sk}}
\newcommand{\pk}{\mathtt{pk}}
\newcommand{\Enc}{\ifmmode\mathsf{Enc}\else\textsf{Enc}\fi}
\newcommand{\Dec}{\ifmmode\mathsf{Dec}\else\textsf{Dec}\fi}
\newcommand{\Ecd}{\ifmmode\mathsf{Ecd}\else\textsf{Ecd}\fi}
\newcommand{\Dcd}{\ifmmode\mathsf{DcdVrfy}\else\textsf{DcdVrfy}\fi}
\newcommand{\Eval}{\ifmmode\mathsf{Eval}\else\textsf{Eval}\fi}
\newcommand{\rot}{\mathsf{rot}}

\renewcommand{\boldsymbol}[1]{#1}

\newcommand{\bincoeff}[2]{\left(\begin{array}{c}#1\\#2\end{array}\right)}

\DeclareMathOperator{\kron}{kron}
\newcommand\blfootnote[1]{%
  \begingroup
  \renewcommand\thefootnote{}\footnote{#1}%
  \addtocounter{footnote}{-1}%
  \endgroup
}

\title{\vspace{-2mm}\bf On the (non-)resilience of encrypted controllers to covert attacks}
\author{Philipp Binfet \and Janis Adamek \and Moritz Schulze Darup}
\date{}

\begin{document}

\maketitle

\textbf{\textit{Abstract}.} {\bf
    \input{abstract}
}

\input{mainmatter}

\clearpage

\appendix
\crefalias{section}{appendix}

\titleformat{\section}{\centering\normalfont\scshape}{Appendix \Alph{section}.}{5pt}{}
\titleformat{\subsection}{\normalfont\it}{\Alph{section}.\arabic{subsection}}{5pt}{}
\titleformat{\subsubsection}{\normalfont\it}{\Alph{section}.\arabic{subsection}.\arabic{subsubsection}}{5pt}{}

\input{appendix}

\end{document}

%% file: abstract.tex
The security of networked control systems (NCS) is receiving increasing attention from both cyber-security and system-theoretic perspectives.
The former focuses on classical IT security goals such as confidentiality, integrity, and availability of process data, while the latter investigates tailored attacks (and detection schemes), including covert and zero-dynamics attacks.
Confidentiality in control systems can, for instance, be achieved by securely outsourcing the evaluation of the controller to third-party platforms, such as cloud services.
The underlying technology enabling such secure computation often is homomorphic encryption (HE).

Recent works in encrypted control have proposed modifications to underlying HE schemes to achieve not only confidentiality but also resilience to certain types of integrity attacks.
While extensions in this direction are
desirable in principle,
we show that
the integrity problem in encrypted control cannot be solved by public-key HE schemes alone due to their inherent malleability.
In other words, the same homomorphisms that enable encrypted control %
in the first place
can be leveraged not only constructively but also destructively.
More precisely, we demonstrate that
NCS are vulnerable to covert attacks,
even when encrypted control is employed.
Remarkably, this remains possible without knowledge of
an unencrypted
model.

Yet, resilience to such attacks can still be achieved through complementary techniques.
We present an approach based on verifiable computation
that integrates with modern homomorphic cryptosystems and
is asymptotically secure
while incurring no communication overhead.

%% file: mainmatter.tex
\section{Introduction and problem statement}
\blfootnote{%
$\copyright$ the authors.
This is an extended version of our paper accepted for presentation at the IFAC World Congress 2026.
This work has been accepted to IFAC for publication under a Creative Commons Licence CC-BY-NC-ND.}
\blfootnote{This work was funded by the Deutsche Forschungsgemeinschaft (DFG, German Research Foundation) –  Grants 422262716 and 503491151.}
Within the area of security of cyber-physical systems,
homomorphic
encryption (HE) has been employed successfully to realize a wide range of control applications in a private way %
by securely outsourcing computation to one or more servers (see \citet{Darup2021Survey} and \citet{schluter2023brief} for an overview).
While HE is suited to address threats against \textit{data confidentiality}, %
it is also well-known to be \textit{malleable} by construction and, hence, cannot satisfy a stronger notion of security which includes \textit{data integrity}~\citep[Ch.~11.2.3]{Katz2014}. %

As a subset of potential threats against cyber-physical systems, covert attacks~\citep{teixeira2012attackModels} stand out as being particularly geared towards networked control systems (NCS)
by manipulating the
I/O data exchanged
between the plant and its remote controller in a coordinated manner in order to stay undetected.
This is usually done by injecting additive biases into these signals during transmission.
More specifically, the control input signal received by the plant is manipulated to disturb the intended system behavior while carefully canceling out the effect on the output measurements visible to the remote controller.
In this way, the attack remains perfectly stealthy while, in principle, arbitrary behavior can be forced upon the system under control.
Unfortunately, this remains true even in the presence of HE due to its inherent malleability.

Recognizing this limitation, attempts have been made to construct HE schemes with built-in resilience to unauthorized modifications of ciphertexts. \citet{fauser2020,fauser2021,fauser2024}, for example, %
describe a modified version of a cryptosystem introduced by~\citet{dyer2019}.
The modified scheme is homomorphic with respect to matrix-vector multiplication and is designed to provide resilience to additive---and, hence, covert---attacks.
However, their construction is limited in that its resilience range depends on the value of the plaintext being encrypted and on the particular realization of the noise terms that are injected into the ciphertext during encryption to provide security.
As a result, resilience is only guaranteed if the attack values are smaller than some threshold.
This essentially limits the scheme's effectiveness, as a covert attacker is free to choose and apply values outside of the resilience range.

A different stream of research is based on the observation that the problems of confidentiality and integrity are essentially orthogonal, allowing for solutions addressing each problem separately.
A system-theoretic approach to integrity is to employ anomaly detectors~\citep{giraldo2018survey}.
In this spirit, \citet{alexandru2022anomaly} pair the outsourced controller with a securely evaluated anomaly detector, both placed at the remote side, and transmit an additional alarm signal back to the plant.
While this approach is suited to address threats such as sensor faults or simple false data injection, %
it suffers from the same malleability issue when it comes to covert attacks due to the detection mechanism being placed at the remote side.

Another approach %
from cryptography, which does not fully rely on the remote side to enforce data integrity
is to attach a message authentication code (MAC) to each message.
MACs ensure that any alteration of an in‑flight message by an unauthorized party is detected
by the recipient
\citep[Ch.~4]{Katz2014}.
There even exist constructions that combine MACs with HE~\citep[][for example]{catalano2013homomorphicMAC} and which could therefore be applied to encrypted control loops.
However, MACs
generally
incur the client-side computational overhead of generating a MAC tag for each message and increase the amount of data that has to be transmitted.
Other approaches focus on verifiable
computation
techniques, which are typically used in the slightly different context of a malicious server but are nevertheless relevant to the topic of covert attacks. %
As a control-oriented example, \citet{stabile2024verifyable} enable verification
by sending additional queries containing ``decoy'' values to the server that evaluates the controller using some HE scheme. %
The responses to these queries are then used to
decide
whether the server's response to the actual payload query should be trusted or not.
The main downside here is the communication overhead incurred by the dummy queries.
A more general and cryptographically rigorous approach is taken by \citet{chatel2024veritas};
they propose special plaintext encoders that can be integrated into modern HE schemes to enable verifiable execution of general homomorphic programs.
Their replication-based encoder, in particular, is similar to our approach in terms of the combination of replication, insertion of challenge values, and permutation.
However, their method incurs linear overhead (in terms of both communication and computation) in the security parameter, which is likely due to its generality.

Our contribution presented in this paper is three-fold:
\begin{itemize}
    \item
        We show that %
        encrypted control loops relying on HE are fundamentally vulnerable to covert attacks.
        Moreover, we describe two %
        instantiations of the homomorphic covert attack and propose a corresponding attack space for classifying such variants (\Cref{sec:covert-attacks-encrypted}).
    \item
        We design
        a
        verifiable computation scheme that integrates with state-of-the-art homomorphic cryptosystems, incurs no communication overhead, and shifts all computational overhead to the server (\Cref{sec:verifiable-encrypted-control}).
    \item We validate our findings in a numerical case study (\Cref{sec:case-study}).
\end{itemize}
\textbf{\textsf{Notation.}}
The identity matrix in $\R^d$ is $\Ib_d$.
The Kronecker product of two matrices $S$ and $T$ is denoted by $\kron(S,T)$.

\section{Preliminaries}

\begin{figure}[bt]
    \centering
    \includegraphics[width=\columnwidth]{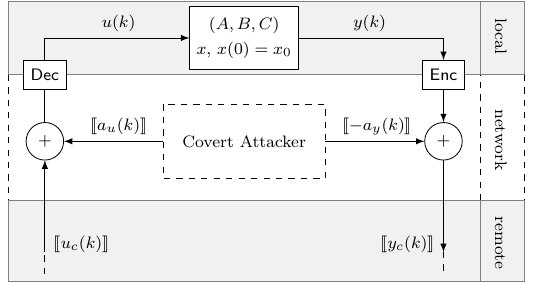}
    \caption{
    Covert attack strategy using the homomorphisms of an encrypted NCS.
    }
    \label{fig:setup-encrypted}
\end{figure}

\subsection{Specification of covert attacks}\label{sec:covert-attacks-basic-spec}

We consider covert attacks on linear systems as specified by~\citet{smith2011appropriating} %
and as illustrated in \Cref{fig:setup-encrypted}.
The underlying idea is simple:
The attacker modifies the desired control input $u_c(k)$ via an additive attack signal $a_u(k)$ resulting in the actual
system input
\begin{equation}\label{eq:input-attacked}
u(k):=u_c(k)+a_u(k).
\end{equation}
To compensate for the effect of the malicious input modification, the attacker also modifies the actual system output $y(k)$ by subtracting an attack signal $a_y(k)$ in such a way that the modified output
\begin{equation}\label{eq:output-attacked}
y_c(k):=y(k)-a_y(k)
\end{equation}
behaves as the actual system output $y(k)$ would
in the attack-free case, where
$u(k)=u_c(k)$.
Specifying the attack is straightforward for linear system dynamics of the form
\begin{align*}
    x(k+1) &= A x(k)+B u(k), \quad x(0):=x_0 \\
    y(k)   &= C x(k).
\end{align*}
The
unmodified system output at time step $k$ then is
\[
y_0(k):=C \left( A^k x_0 + \sum_{i=0}^{k-1} A^{k-i-1} B u_c(i) \right).
\]
Now, assuming the attacker applies the input attack signal $a_u(\cdot)$, the output of the attacked system becomes
\[
y(k)=y_0(k)+C \sum_{i=0}^{k-1} A^{k-i-1} B a_u(i).
\]
Hence, the output attack needs to compensate for the malicious forced motion component.
Under the additional assumption that the attacker has access to a model of the system in form of the matrices $A$, $B$, and $C$, this is obviously achieved by choosing
\[
a_y(k):=C \sum_{i=0}^{k-1} A^{k-i-1} B a_u(i).
\]
Remarkably, the same output attack sequence can also be computed recursively using the following dynamics:
\begin{subequations}\label{eq:delta-system}
\begin{align}
\label{eq:delta-dynamics}
    \Dx(k+1) &= A \Dx(k)+B a_u(k), \quad \Dx(0):=0 \\
\label{eq:output-attack-from-delta-dynamics}
    a_y(k)   &= C \Dx(k).
\end{align}
\end{subequations}
Without loss of generality, we assume that the attack starts at $k=0$.

\subsection{Finite-length covert attacks}\label{sec:covert-attacks-finite-length}
While the specification of $a_y(k)$ is straightforward in principle, an actual attack should consider an additional feature.
In fact, even if $a_u(k)$ is chosen to be non-zero only for $k$ smaller than a finite $k_u \in \N$, $a_y(k)$ may be non-zero for every $k \in \N$.
In other words, a perfectly stealthy attack might require the attacker's ``attention'' for an indefinite amount of time.
Clearly, finite attention is desirable in practice.
Two basic approaches to achieving this are discussed in the related literature.
The first approach assumes a Schur stable $A$ and exploits the fact that commonly employed anomaly detectors operate on a non-zero detection threshold, below which no alarm will be raised.
Consequently, $a_u(k)=0$ for every $k \geq k_u$ will eventually lead to $\|a_y(k)\|<\epsilon$ for all $k\geq k_y$ for some $k_y\in\N$ and some $\epsilon>0$ corresponding to the detection threshold.
Put differently, the active attack phase ($k<k_u$) here is followed by a passive \textit{cooldown} phase ($k_u\leq k < k_y$) after which
output compensation via~\eqref{eq:output-attack-from-delta-dynamics} can be terminated without sacrificing post-attack stealthiness.
The second approach~\citep{gheitasi2020finite} employs an active strategy to achieve a specified cooldown duration and to provide perfect stealthiness ($a_y(k)=0$) for every $k$ after the attack.
In order to do so, special input attack signals are required during cooldown and the pair $(A,B)$ must be controllable.
One can then consider an input attack of length
$L \geq n$
and design the last $\Dk \in \N \cap [n,L]$ non-zero input modifications $a_u(\cdot)$ such that $\Dx(L)=0$.
Clearly, this implies $\Dx(k)=0$ and $a_y(k)=0$ for every $k\geq L$.
$\Dk$ here denotes the length of the active cooldown phase.
\citet{gheitasi2020finite} proposed an optimization-based selection of the
attack sequence
$a_u(L-\Dk),\dots,a_u(L-1)$, which allows for additional features such as minimal control effort or constraint satisfaction~\citep{gheitasi2022finiteCovertConstrained}.
Here, we limit the discussion to the basic form of the attack and focus on the special case $\Dk:=n$ to avoid the need for explicit optimization,
resulting in
\begin{equation}\label{eq:cooldown-sequence}
    \begin{bmatrix}
        a_u(L-n) \\
        \vdots \\
        a_u(L-1)
    \end{bmatrix}
    =
    \Cc_n^\dag A^n \Dx(L - n),
\end{equation}
where $\Cc^\dag$ denotes the Moore-Penrose inverse of the controllability matrix
\(
    \Cc_n = \begin{bmatrix}
        B & AB & \dots & A^{n-1}B
    \end{bmatrix}.
\)

\subsection{Cryptographic basics and homomorphic primitives}\label{sec:HE-basics}

Broadly speaking, a public-key cryptosystem provides two mappings $\Enc_{\pk}: \ptSpace\to\ctSpace$ and $\Dec_{\sk}: \ctSpace\to\ptSpace$ between the space of messages or plaintexts, $\ptSpace$ (typically a finite integer set), and the space of encrypted messages or ciphertexts, $\ctSpace$.
The encryption procedure is parameterized on a public key $\pk$ and, in principle, available to anyone, while decryption requires a secret key $\sk$ and, thus, can only be carried out by authorized parties.
We will write \Enc{} and \Dec{} in the following without explicitly denoting the corresponding keys unless the context requires disambiguation.
A ciphertext which corresponds to the plaintext $m\in\ptSpace$ will be denoted by $\cipher{m}\in\ctSpace$, i.e., it satisfies $\Dec(\cipher{m})=m$.%

Homomorphic encryption (HE) refers to cryptosystems under which, in addition to the encryption-decryption mechanism, it is possible to manipulate ciphertexts in a systematic way such that the corresponding plaintext, after decryption, reflects the result of a particular (arithmetic) operation performed on it. More specifically, given an operation $\star: \ptSpace\times\ptSpace\to\ptSpace$ in the plaintext space and a corresponding operation $\ast: \ctSpace\times\ctSpace\to\ctSpace$ in the ciphertext space, then \Dec{} is a (group) homomorphism from $(\ctSpace, \ast)$ to $(\ptSpace, \star)$ if
\begin{equation}\label{eq:homomorphism-generic}
    \Dec(a \ast b) = \Dec(a) \star \Dec(b)
\end{equation}
for every $a,b\in\ctSpace$.
In other words, the plaintext operation $\star$ can be evaluated in an equivalent way using $\ast$ in the encrypted domain.
Common examples for %
$\star$ are addition and multiplication on $\ptSpace$.
In this paper, we will refer to their counterparts on $\ctSpace$ as \emph{homomorphic addition} ($\oplus$) and \emph{homomorphic multiplication} ($\odot$), respectively.
By extension, we use $\Eval_h$ to denote the \emph{homomorphic evaluation of a function $h$} such that $\Dec(\Eval_h(\cipher{\cdot}))=h(\cdot)$.

\begin{rem}[Advanced homomorphic primitives]\label{rem:advanced-HE}
    ~~Some state-of-the-art arithmetic HE schemes support vectorization or \emph{single instruction multiple data} (SIMD) through \emph{ciphertext packing}, meaning that a single ciphertext can represent an entire vector %
    in the form of multiple plaintext slots;
    the homomorphic operations in this case process all %
    slots
    simultaneously.
    Furthermore, the palette of supported homomorphic operations typically also includes
    \begin{enumerate}
        \item efficient analogous versions of $\oplus$ and $\odot$ in which one of the operands is a plaintext;
        \item additional binary operations such as subtraction; %
        \item additional unary operations such as negation and circular rotation of the
        plaintext slots
        (an analogous unary version of~\eqref{eq:homomorphism-generic} holds in this case).
    \end{enumerate}
    We will collect these additional capabilities under the term \emph{advanced HE}.
\end{rem}

Besides the mere correctness of their functionality, cryptosystems (homomorphic or otherwise) are only as good as the guarantees they can provide regarding their ability to protect the confidentiality of the handled data.
And in fact, it is standard practice in modern cryptography to prove the security of cryptosystems in a mathematically rigorous way.
Homomorphic cryptosystems are no different in this regard; many HE schemes exist that are provably secure under standard cryptographic assumptions (typically regarding the hardness of certain mathematical problems).
Practically speaking, this means that
HE is suited to provide data \emph{confidentiality} by virtue of \Dec{} being intractable to evaluate when $\sk$ is not known.
However, HE on its own cannot provide data \emph{integrity}, since ciphertexts can readily be modified through homomorphic operations, which means that HE is \emph{malleable} by construction.
This malleability of HE is
essentially %
what enables the type of attacks described in \Cref{sec:covert-attacks-basic-spec} to be launched even when HE is used to keep the transmitted data confidential, which will be discussed in \Cref{sec:covert-attacks-encrypted}. %

\begin{rem}[Ciphertext notation for higher-level objects]
    The notation $\cipher{\cdot}$ was previously introduced only for encryptions of individual elements from the message space.
    For the sake of readability, we will use the same notation to also denote encryptions of vectors and matrices in the following.
    Hints regarding a possible concrete implementation
    are given in \Cref{sec:enc-vector-matrix-repr-teaser}.
\end{rem}

\subsection{Homomorphic algorithms}\label{sec:HE-algs}
The basic homomorphic operations described above can be used to define more complex algorithms.
Two such algorithms, which will be of importance in this paper, are encrypted matrix multiplication $\cipher{ST} := \cipher{S}\otimes\cipher{T}$ and matrix-vector multiplication $\cipher{Sv} := \cipher{S}\otimes\cipher{v}$ for a vector $v$ and matrices $S,T$ of appropriate size.
While it is clearly possible to define $\otimes$ in terms of just $\oplus$ and $\odot$, more efficient realizations using advanced HE exist; the implementation used in this paper is described in \Cref{app:enc-matrix-vector-operations}.
Furthermore, it is possible to define an algorithm for finding the Moore-Penrose inverse of an encrypted matrix based on encrypted matrix multiplication.
Such an algorithm can be used to solve linear least-squares problems on encrypted data; notably, this class of problems includes the application of system identification on encrypted I/O data.
We refer the interested reader to~\citet{adamek2024sysid} for details.

\section{Homomorphic covert attacks}\label{sec:covert-attacks-encrypted}
We shall now introduce and analyze two variants of the covert attacks described in \Cref{sec:covert-attacks-basic-spec} that operate on encrypted I/O data.
The two cases will differ in terms of the system knowledge and the homomorphic capabilities granted to the attacker.

\subsection{Plaintext model and additive HE}\label{sec:attack-with-plain-model}
In the first scenario, the attacker is assumed to have access to $(A,B,C)$ in unencrypted form, for example from expert knowledge or because the system under attack is a common appliance for which a suitable model is publicly available.
After choosing $a_u(k)$, the attacker can then readily evaluate~\eqref{eq:delta-system}
and~\eqref{eq:cooldown-sequence} in plaintext for all $k$.
The attack signals are injected into the input and output channel, respectively, via
\begin{align*}
    \cipher{u(k)}   &= \cipher{u_c(k)} \oplus \Enc(a_u(k)), \\
    \cipher{y_c(k)} &= \cipher{y(k)} \oplus \Enc(-a_y(k))
\end{align*}
according to \eqref{eq:input-attacked}--\eqref{eq:output-attacked}.
Note that $\oplus$ is the only homomorphic operation the attacker needs to
implement the attack.

\subsection{Encrypted model and advanced HE}\label{sec:attack-with-enc-model}

In the second scenario,
the attacker's knowledge of the system model is reduced to $(\cipher{A}, \cipher{B}, \cipher{C})$.
A possible way that the encrypted model might have been obtained is through encrypted system identification on intercepted encrypted I/O data (see \Cref{sec:HE-algs}).

While evaluating \eqref{eq:input-attacked}--\eqref{eq:output-attacked} still only requires $\oplus$, this is not the case anymore for~\eqref{eq:delta-system}
or \eqref{eq:cooldown-sequence}.
Instead, the fully encrypted equivalents of these expressions,
\begin{subequations}
\begin{align}
\label{eq:delta-dynamics-encrypted}
    \cipher{\Dx(k+1)} &= \cipher{A}\otimes\cipher{\Dx(k)} \oplus\cipher{B}\otimes\Enc(a_u(k)),\\
    \cipher{a_y(k)}   &= \cipher{C}\otimes \cipher{\Dx(k)}
\end{align}
\end{subequations}
and
\begin{equation}%
    \begin{bmatrix}
        \cipher{a_u(L-n)} \\
        \vdots \\
        \cipher{a_u(L-1)}
    \end{bmatrix}
    =
    \cipher{\Cc_n^\dag}\otimes\left(\cipher{A^n}\otimes\cipher{\Dx(L - n)}\right),
\end{equation}
must be evaluated, which requires advanced HE capabilities (see Remark~\ref{rem:advanced-HE}).
Furthermore, due to the recursive nature of~\eqref{eq:delta-dynamics} resp.~\eqref{eq:delta-dynamics-encrypted}, the depth of the computational circuit increases with the attack duration (which is, however, finite; see \Cref{sec:covert-attacks-finite-length}).
Similarly, a moderately deep circuit is required for computing $\cipher{\Cc^\dag_n}$.

\subsection{Attack classification}
The two variants of the homomorphic covert attack described above differ primarily in two aspects: System knowledge and homomorphic capability.

\begin{figure}
    \centering
    \includegraphics[scale=1.0]
    {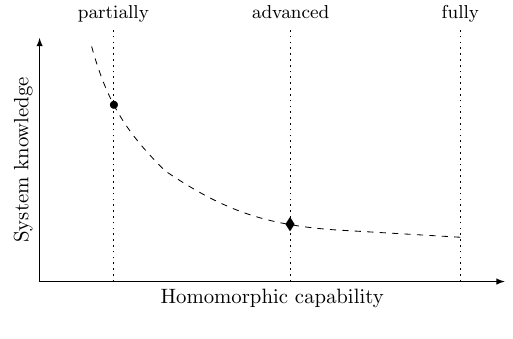}
    \caption{%
        Visualization of the homomorphic covert attack space.
        The two marked points correspond to the attack variants described in Sects.~\ref{sec:attack-with-plain-model} ($\bullet$) and~\ref{sec:attack-with-enc-model} ($\blacklozenge$).
        The dashed line
        denotes
        attack variants with comparable tradeoffs between the two dimensions and, hence, similar overall resource requirements.
    }
    \label{fig:attack-space}
\end{figure}

\emph{System knowledge} is the
information about the system under attack that is available to the attacker \citep{teixeira2012attackModels}.
It may vary in quantity and quality, where quality may refer to the precision, utility, or severity of the data in question.
For example, $\cipher{A}$ is considered to have lower severity than $A$, as it contains less immediately usable information.
\emph{Homomorphic capability} refers to the type and amount of homomorphic operations that can be utilized by the attacker.
For the purposes of this paper, we define the following characteristic points along this dimension:
partially HE (either $\oplus$ or $\odot$ is available), advanced HE (see Remark~\ref{rem:advanced-HE}), and fully HE.
The latter refers to the ability to evaluate arbitrary-depth computational circuits, which is typically achieved via so-called bootstrapping.
Fully HE is not required in this paper and only mentioned to provide a reasonable upper bound of the homomorphic capability axis in the attack space.
See \citet{armknecht2015guideFHE} or \cite{marcolla2022surveyFHE} for an overview.
We propose an attack space, spanned by these two dimensions, along which homomorphic covert attack variants may be classified.
A qualitative illustration is given in \Cref{fig:attack-space}.

In this context, the first attack variant can be placed in the upper-left region of the attack space in~\Cref{fig:attack-space}, because it can be implemented with only very basic homomorphic operations, but only if full system knowledge in terms of an unencrypted model is available.
By only $\oplus$ being required, this attack can be applied for a wide range of homomorphic cryptosystems.

The second attack variant, in contrast, constitutes the opposite extreme being located towards the lower-right region of the attack space.
The necessary side information is arguably less severe here since the system model only needs to be provided in an encrypted form.
While this makes the attack more likely in principle, it also demands considerably more complex homomorphic functionality.

In summary, it appears that system knowledge and homomorphic capability are, in a sense, competing dimensions;
the tradeoff between the two can be understood from two perspectives:
For a fixed set of HE capabilities, system knowledge of a certain quantity and quality must be available to enable the attack.
Conversely, for a given amount of system knowledge of given quality, a certain minimum level of HE capability is necessary to successfully launch the attack.

\section{Towards zero-overhead verifiable encrypted control}\label{sec:verifiable-encrypted-control}
As has been demonstrated above, control loops featuring remote controller evaluation using HE alone are fundamentally vulnerable to homomorphic covert attacks.
To address this problem, we propose a tailored verifiable computation scheme which improves on existing approaches in terms of security, flexibility, and efficiency.

\subsection{Motivation}
As discussed in Remark~\ref{rem:advanced-HE}, modern homomorphic cryptosystems support packing multiple plaintext slots into a single ciphertext.
Typical parameter choices\footnote{%
For schemes derived from the ring learning with errors problem, the key length is the decisive factor in this regard.}
lead to instantiations supporting rather large plaintext slot counts ranging from $2^{12}$ to $2^{16}$.
It is hence not uncommon that many of the available slots are not used.
One key element of our approach is to populate (a portion of) these unused slots to enable the simultaneous processing of the regular payload data and challenge data, which will then be used for verification against precomputed values.
In general, this will require a modification of the function $h$ before uploading it to the server.
Before describing our approach in detail, we make the following simplifying assumption:
\begin{assum}\label{assum:outsourced-function-domain-codomain-dimensions}
    The function to be outsourced is of the form $h:\R^d\to\R^d$.
\end{assum}
Note that, without loss of generality, Assumption~\ref{assum:outsourced-function-domain-codomain-dimensions} can be satisfied by applying suitable zero padding.

\subsection{A lightweight verifiable computation scheme}\label{sec:verifiable-computation-scheme}
Our verification scheme can be formalized as the collection of the client-side algorithms \textsf{Setup}, \Ecd{}, and \Dcd{}, which are specified below.
\newcommand{\verin}{w}            %
\newcommand{\verout}{z}           %
\newcommand{\varverout}{\zeta}    %
\begin{description}
    \item[\textsf{Setup}($N, d, h$)]
        Choose an even expansion factor $\lambda\in 2\N$ such that $\lambda d\leq N/2$ and a detection threshold $\varepsilon > 0$.
        Generate $M\in\N$ \emph{challenge values} $c_i$, precompute $h(c_i)$ and store $(c_i, h(c_i))$, $1\leq i\leq M$, at the client.
        Transmit the lifted function $\tilde h:\R^{\lambda d}\to\R^{\lambda d}$, $\tilde\verin =(\verin_1, \dots, \verin_\lambda)\mapsto (h(\verin_1), \dots, h(\verin_\lambda))$ to the server.
    \item[\Ecd($\verin$)]
        On input $\verin\in\R^d$, generate a uniformly random permutation matrix $P^\prime\in\{0, 1\}^{\lambda\times\lambda}$ and
        store
        the block-wise permutation matrix $P:=\kron(P^\prime, \Ib_d)$.
        Draw %
        integers $i_1, i_2, \dots, i_{\lambda/2}$ uniformly at random from $[1, M]$ and select the challenge values $c_{i_r}$, $1\leq r\leq \lambda/2$.
        Form the augmented input by concatenating  $\lambda/2$ replicas of the payload and the selected challenge values as $\verin^\prime := [\verin^\top, \dots, \verin^\top, c_{i_1}^\top, \dots, c_{i_{\lambda/2}}^\top]^\top\in\R^{\lambda d}$ and output the block-shuffled version $\tilde{\verin}:=P\verin^\prime$.
    \item[\Dcd{}($\tilde{\verout}$)]  %
        On input $\tilde{\verout}\in\R^{\lambda d}$, parse the blocks of $\verout^\prime:=P^{-1}\tilde{\verout}$ as $(\varverout_1, \cdots, \varverout_{\lambda/2}, \varsigma_1, \dots, \varsigma_{\lambda/2})$.
        Verify that $\|\varsigma_r - h(c_{i_r})\|_\infty\leq\varepsilon$ for all $1\leq r\leq \lambda/2$ using the precomputed values.
        If verification passes, conclude that $\varverout_i\approx h(\verin)$ for all $1\leq i\leq \lambda/2$ with high probability and output
        one of the $\varverout_i$ uniformly at random.
        Otherwise, %
        output the error symbol $\perp$ and
        terminate the interaction with the server.
\end{description}
\begin{figure}
    \centering
    \includegraphics[scale=1]{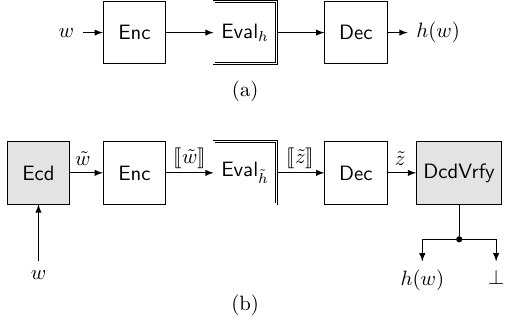}
    \caption{%
        Secure computation pipelines for evaluating $h$, without (a) and with (b) verification.
        Double‑outlined boxes denote server-side operations.
    }
    \label{fig:HE-pipeline-with-and-without-verification}
\end{figure}

To make the above verification scheme compatible with an existing HE pipeline for evaluating $h$ (see \Cref{fig:HE-pipeline-with-and-without-verification}a), \Ecd{} and \Dcd{} are added as pre- and postprocessing steps, respectively (see \Cref{fig:HE-pipeline-with-and-without-verification}b).
Accordingly, the server must be set up to evaluate the lifted function $\tilde h$ instead of $h$.
How this can be done efficiently is shown
in detail in \Cref{sec:lifting}
for the example of linear or affine functions.
The discussion there also addresses the use of SIMD to minimize computational overhead.

\begin{rem}[Detection threshold]
    \Dcd{} implements an approximate detection criterion based on the threshold $\varepsilon$ to account for small deviations
    that
    occur naturally during the evaluation of $\tilde{h}$ due to finite-precision arithmetic.
\end{rem}

\subsection{Attack model}\label{sec:attack-model}
In our model, the attacker's malicious activity is confined to the modification of the measurements $\cipher{y(k)}$ and control actions $\cipher{u_c(k)}$ according to \Cref{fig:setup-encrypted}.
The attacker's goal is to
alter these values systematically (see \Cref{sec:covert-attacks-encrypted}) while staying stealthy.
Stealthiness requires that all $\varsigma_{i_r}$ should be intact upon decoding.
To meet this requirement,
the attacker must modify all $\lambda/2$ blocks carrying the replicas of the payload consistently and simultaneously (and leave the remaining $\lambda/2$ blocks untouched).
We ascribe \emph{instantaneous success} to the attack if this is the case in a particular time step.

Since the permutation $P$ is unknown to the attacker, we assume the attacker guesses the indices of these blocks uniformly at random.
The resulting attack success probability (resp.~the scheme's attack detection rate) is derived next.

\subsection{Attack success probability and detection rate}\label{sec:detection-rate}
The attacker guesses $\lambda/2$ out of $\lambda$ block indices.
In a given time step, this guess is successful with probability
\begin{equation}\label{eq:attack-success-probability-instantaneous}
    p_\text{succ}(1)
    =\bincoeff{\lambda}{\lambda/2}^{-1}
    = \frac{\frac{\lambda}{2}! (\lambda -\frac{\lambda}{2})!}{\lambda !}.
\end{equation}
It can be shown that $p_\text{succ}(1)\leq 2^{-\lambda/2}$ for $\lambda>1$, i.e., the \emph{instantaneous attack success probability} is a negligible function in $\lambda$ (see \Cref{app:negligible-attack-success-probability}).
Consequently, the verification scheme is asymptotically secure with security parameter $\lambda$.

Remaining stealthy over an attack of length $L\geq 1$ time steps is possible only with probability
\begin{equation}\label{eq:attack-success-probability}
    p_\text{succ}(L)={p_\text{succ}(1)}^L,
\end{equation}
since the permutation $P$ is resampled in every time step.
This results in an improved concrete upper bound for the \emph{cumulative attack success probability} of $p_\text{succ}(L)\leq 2^{-L\lambda/2}$.
Conversely, an attack of length at least $L$ is detected with probability at least $1-p_\text{succ}(L) \geq 1 - 2^{-L\lambda/2}$.
For example, for $L=10$, a rather small expansion factor of $\lambda=4$ already yields 40-bit security or, equivalently, a cumulative attack success probability of less than $10^{-12}$.

\section{Case study and implementation details}\label{sec:case-study}
We will now validate both the feasibility of homomorphic covert attacks and the effectiveness of our verifiable computation approach on two
instantiations of the homomorphic covert attack
in a numerical case study  %
and discuss relevant implementation details.

\subsection{System model}
We demonstrate the viability of the homomorphic covert attacks from \Cref{sec:covert-attacks-encrypted} on the model
\begin{gather*}
        A= \begin{bmatrix}
               0.984 & 0.000 & 0.041 & 0.000\\
               0.000 & 0.989 & 0.000 & 0.033\\
               0.000 & 0.000 & 0.959 & 0.000\\
               0.000 & 0.000 & 0.000 & 0.967
           \end{bmatrix},\;
        B= \begin{bmatrix}
              0.083 & 0.001\\
              0.001 & 0.063\\
              0.000 & 0.047\\
              0.031 & 0.000
           \end{bmatrix},\\
        C= \begin{bmatrix}
              0.500 & 0.000 & 0.000 & 0.000\\
              0.000 & 0.500 & 0.000 & 0.000
           \end{bmatrix},
\end{gather*}
which is a linearized version of the quadruple-tank process~\citep{johansson2000quadtank}.
For simplicity and since our focus in this paper is on linear systems, we use this linearized dynamics not only for the design of the controller and the attacks but also to simulate the actual process behavior.
As stated in~\Cref{sec:covert-attacks-basic-spec}, we define $k=0$ as the starting point of the attack. Hence, the system is initialized with $x(-20)=[\begin{array}{cccc} 1&1&0&0 \end{array}]^\top$ and is driven to the setpoint $x_\text{ref} = [\begin{array}{cccc}1.15&1.20&0.17&0.13\end{array}]^\top$
using the static output feedback controller
\begin{equation}\label{eq:output-feedback-control-law}
    u_c(k) =
    -\begin{bmatrix}
      11.545 & 0.061\\
      1.609 & 11.131
    \end{bmatrix} y_c(k) + u_0,
\end{equation}
where $u_0 = [\begin{array}{cc}6.80&7.76\end{array}]^\top$.
\subsection{Cryptosystem}\label{sec:HE-instantiation-and-linalg} %
We choose the
lattice-based homomorphic CKKS cryptosystem~\citep{cheon2017ckks}
that is especially suitable for encrypted control.
This is due to its ability to handle arithmetic calculations over real numbers with high accuracy, its support for advanced homomorphic primitives (see Remark~\ref{rem:advanced-HE}), the natural SIMD support, and the state-of-the-art performance measures. We instantiate the cryptosystem with a ring dimension (i.e., the key length) of $N=2^{17}$ and a scaling value of $2^{60}$, which are typical values in encrypted control systems.

\subsection{Encrypted vector and matrix representation}\label{sec:enc-vector-matrix-repr-teaser}
We leverage ciphertext packing to represent a plaintext vector $v$ by the single ciphertext $\cipher{v}$.
If the encryption of a matrix $S$, $\cipher{S}$, is accordingly defined as a collection of its encrypted diagonals, the matrix-vector multiplication algorithm of \citet{halevi2014} can be readily used.
Their ideas can then be extended to matrix-matrix multiplication in a straightforward fashion.
See \Cref{app:enc-matrix-vector-operations} for implementation details.

\subsection{Lifting of linear and affine functions}\label{sec:lifting}
As described in \Cref{sec:verifiable-computation-scheme}, the outsourced function must be modified to be compatible with the augmented inputs and outputs in the verification scheme.
From a syntactic point of view, the lifted function $\tilde h$ is only required to evaluate the original function $h$ on each block of its input and output the result blocks, concatenated in order.
From an efficiency point of view, doing so explicitly should in general be avoided, since this would incur a $\lambda$-times increase in computation time;
instead, one should use SIMD to evaluate all blocks simultaneously.

For affine functions as in~\eqref{eq:output-feedback-control-law}, this can be achieved as follows.
Let $h(w) = K w + v$.
Define
\(
    \bar{K} = \begin{bmatrix} K & I \end{bmatrix}
\)
and
\(
    \bar{w} = \begin{bmatrix} w^\top & v^\top \end{bmatrix}^\top
\)
to obtain a linear description of the original function in terms of
\(
\bar h(\bar w) = \bar{K} \bar{w}.
\)
The lifted function can then can be expressed as
\(
    \tilde h (\tilde w) = \tilde K \otimes \tilde w,
\)
where
\(
    \tilde K = \kron(I_\lambda, \bar{K})
\)
and
\(
    \tilde w = \kron(I_\lambda, \bar{K}).
\)

Consequently, the encrypted implementation comes down to a single encrypted matrix-vector multiplication, i.e.,
\(
\Eval_{\tilde h}(\cdot) = \cipher{\tilde K} \otimes (\cdot),
\)
which can be implemented efficiently.
Most notably, $\Eval_{\tilde h}$ can make full use of SIMD through the vectorized matrix-vector multiplication algorithm and several related optimizations (described in detail in \Cref{app:enc-matrix-vector-operations,app:optimizations}).
In effect, the overhead of evaluating $\Eval_{\tilde h}$ compared to the analogous $\Eval_{h}$ is minimal.

\subsection{Homomorphic covert attacks}
\begin{figure}
    \includegraphics{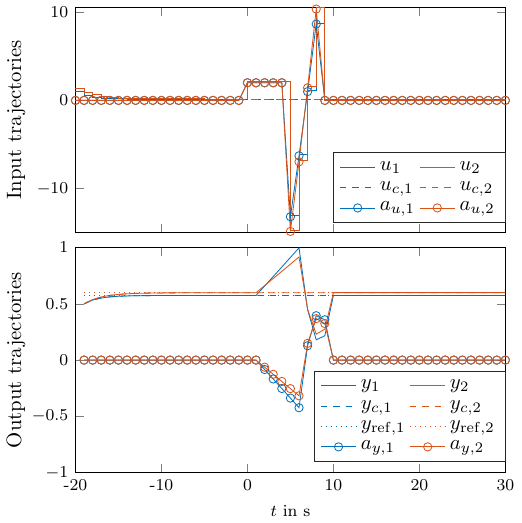}
    \caption{%
        Experimentally obtained I/O trajectories
        in
        both homomorphic covert attack
        variants.
    }
    \label{fig:trajectories-enc}
\end{figure}

We now show
a concrete homomorphic covert
attack with $a_u(k)=\begin{bmatrix}
    2&2
\end{bmatrix}^\top$ for $k\in[0,4]$, which is subsequently terminated with proper cooldown at $k=9$ such that $L=10$. The two scenarios from \Cref{sec:attack-with-plain-model} (plaintext model) and \Cref{sec:attack-with-enc-model} (encrypted model) are both visualized in \Cref{fig:trajectories-enc}, since they share nearly identical trajectories. The figure shows the input and output sequences at both the system and the remote controller before, during, and after the attack. The attack phase is entirely stealthy, as the controller receives the expected input and output sequence and can therefore not identify malicious activity with anomaly detectors over HE.
Once the cooldown phase is concluded, all deviations between $u_c(k)$ and $u(k)$ as well as $y_c(k)$ and $y(k)$ have been eliminated, which preserves stealthiness even after the attack.
Therefore, the covert attack and cooldown can be successfully implemented with both stages of model knowledge. However, the configuration from \Cref{sec:attack-with-enc-model} is more computationally demanding.

\subsection{Verifiable computation approach}

We conduct an empirical study comprised of multiple experiments to validate the concrete effectiveness of our verifiable computation scheme from \Cref{sec:verifiable-computation-scheme}.
In each experiment, we first fix a particular $\lambda\in[2,16]$ and instantiate the scheme.
(setting $N=2^{17}$ as before).
With these choices, we could accommodate a payload dimension $d$ of up to $N/(2\lambda)=2^{12}$ even for the highest value of $\lambda$, which is far beyond what is required in typical control applications.
Then, we simulate an attacker launching a finite-length covert attack of length $L=10$ who is further assumed to follow the strategy outlined in \Cref{sec:attack-model}.
We then repeat this simulation 100,000 times, recording for each run
the number of time steps passed since the beginning of the attack until the attack is detected.

The experimental results reported in \Cref{tab:detection-rate} show that for all $\lambda > 2$ the attack was successfully detected before it could be completed.
Even for $\lambda=2$ (the lowest possible value), the chance of not detecting the attack was lower than $0.1\%$.
Note also that $\lambda=4$ was sufficient to detect the attack in virtually all cases by the 6-th time step.
For $\lambda=16$, $99.99\%$ of the time the attack was detected immediately.
These findings also nicely align with the analytical expression for $p_\text{succ}(L)$ in~\eqref{eq:attack-success-probability}.

\begin{table}[h]
    \centering
    \caption{%
        Experimental detection rates.
        Each row reports the
        percentage of cases
        (sample size: 100,000)
        detected
        in the $k^\ast$-th step of the attack.
        `/' denotes cases in which the attack was not detected within 10 steps.
    }
    \label{tab:detection-rate}
    \begin{tabular}{ccccc}
    \toprule
        $k^\ast$ & $\lambda=2$         & $\lambda=4$         & $\lambda=8$         & $\lambda=16$       \\
     \midrule
    1 & %
    49.98\%%
    & %
    83.43\%%
    & %
    98.58\%%
    & %
    99.99\%%
    \\
    2 & %
    24.84\%%
    & %
    13.80\%%
    & %
    1.40\%%
    & %
    0.01\%%
    \\
    3 & %
    12.53\%%
    & %
    2.27\%%
    & %
    0.02\%%
    & %
    0.00\%%
    \\
    4 & %
    6.43\%%
    & %
    0.42\%%
    & %
    0.00\%%
    & %
    0.00\%%
    \\
    5 & %
    3.08\%%
    & %
    0.07\%%
    & %
    0.00\%%
    & %
    0.00\%%
    \\
    6 & %
    1.59\%%
    & %
    0.01\%%
    & %
    0.00\%%
    & %
    0.00\%%
    \\
    7 & %
    0.80\%%
    & %
    0.00\%%
    & %
    0.00\%%
    & %
    0.00\%%
    \\
    8 & %
    0.39\%%
    & %
    0.00\%%
    & %
    0.00\%%
    & %
    0.00\%%
    \\
    9 & %
    0.18\%%
    & %
    0.00\%%
    & %
    0.00\%%
    & %
    0.00\%%
    \\
   10 & %
   0.08\%%
   & %
   0.00\%%
   & %
   0.00\%%
   & %
   0.00\%%
   \\
   \midrule
   / & 0.09\% &  0.00\% &  0.00\% &  0.00\%
   \\
     \bottomrule
    \end{tabular}
\end{table}

\section{Conclusion and outlook}
In this
work,
we demonstrated that encrypted controllers are vulnerable to covert attacks by the inherent malleability of HE and therefore require verifiable computation schemes.
We introduced an attack space for homomorphic covert attacks
to aid in classifying possible attack scenarios by the tradeoff they make between the two dimensions `system knowledge' and `homomorphic capability'.
As a
countermeasure
to
homomorphic covert attacks,
we designed a verifiable computation scheme with zero communication overhead and asymptotic security, exploiting the SIMD feature of state-of-the-art HE schemes. Furthermore, we described two extreme cases in the attack space and used them to test our verifiable computation method in a numerical case study.

Future research
will be geared towards exploring possible extensions of the verification scheme
in two main directions:
First, we will investigate the possibility of recovering the payload even when an attack is detected. Second, we strive to improve the efficiency of the scheme and, thus, reduce the server-side computational overhead for certain classes of functions $h$.

%% file: appendix.tex
\section{Upper bound for instantaneous attack success probability}\label{app:negligible-attack-success-probability}
We want to show that $p_\text{succ}\leq 2^{-\lambda/2}$ for $\lambda>1$.
From~\eqref{eq:attack-success-probability-instantaneous} we have
\begin{align*}
    p_\text{succ}(1)
    &= \left(\begin{array}{c}
            \lambda \\
            \lambda/2
       \end{array}\right)^{-1}
    = \frac{\frac{\lambda}{2}! \frac{\lambda}{2}!}{\lambda !}
    \\
    &= \frac{\frac{\lambda}{2}\cdot\frac{\lambda-2}{2}\cdot\dots\cdot\frac{\lambda-(\lambda-2)}{2}}{\lambda\cdot(\lambda-1)\cdot\dots\cdot(\lambda - (\frac{\lambda}{2}+1))}
    = \prod_{i=0}^{\frac{\lambda}{2}-1} \frac{\frac{\lambda-2i}{2}}{\lambda-i}
    \\&%
    =2^{-\frac{\lambda}{2}}\prod_{i=0}^{\frac{\lambda}{2}-1} \frac{\lambda-2i}{\lambda-i}.
\end{align*}
Since $\lambda - 2i\leq \lambda-i$ for all $0\leq i<\frac{\lambda}{2}$,
the product in the last line is at most $1$, which concludes the proof.

\section{Essential encrypted matrix and vector operations}\label{app:enc-matrix-vector-operations}
To simplify the discussion, we make the following assumption (w.l.o.g.) about the shapes of vectors and matrices that need to be handled.
\begin{assum}\label{assum:encrypted-dimensions}
    All matrices and vectors subject to encryption are of size $\batchsize\times\batchsize$ and $\batchsize$-dimensional, respectively, where $\batchsize$ is a power of two integer.
\end{assum}
Assumption~\ref{assum:encrypted-dimensions} reflects the fact that CKKS is typically implemented over rings of a power-of-two dimension $N$.
Consequently, when ciphertext packing is used, the subring dimension (sometimes also called batch size) must be a power of two as well~\citep[see][]{cheon2017ckks}.
Practically speaking, this requirement can always be met by applying proper zero-padding.

We represent a vector $v=[v_i]_{0\leq i < \batchsize}\in\R^\batchsize$ by the single ciphertext $\cipher{v}$.
Using the diagonal method~\citep{halevi2014}, a matrix $S=[s_{i,j}]_{0\leq i,j < \batchsize}\in\R^{\batchsize\times\batchsize}$ can be split into (at most) $\batchsize$ vectors $(S_0, S_1, \dots, S_{\batchsize - 1})$, where $S_i:=[s_{j,i+j}]_{0\leq j < \batchsize}\in\R^\batchsize$ refers to the $i$-th \emph{wrapping diagonal} of $S$ and is represented by the ciphertext $\cipher{S_i}$.
We define the matrix ciphertext $\cipher{S}:=(\cipher{S_i})_i$ as the $\batchsize$-tuple of its encrypted diagonals.
Matrix-vector multiplication can then be defined via
\begin{equation}\label{eq:matrix-vector-mult-diagonal}
    \cipher{S v} = \cipher{S}\otimes\cipher{v} := \bigoplus_{i=0}^{\batchsize-1} \cipher{S_i} \otimes \rot_{i}(\cipher{v}).
\end{equation}
This algorithm is straightforwardly extended to matrix-matrix multiplication by substituting $\cipher{v}$ with the appropriate diagonal of the right-hand side operand:
\begin{equation}\label{eq:matrix-matrix-mult-diagonal}
    \cipher{(S T)_k} = (\cipher{S}\otimes\cipher{T})_k := \bigoplus_{i=0}^{\batchsize-1} \cipher{S_i} \otimes \rot_{i}(\cipher{T_{k-i}}).
\end{equation}
In~\eqref{eq:matrix-vector-mult-diagonal}--\eqref{eq:matrix-matrix-mult-diagonal}, $\rot$ refers to ciphertext rotation;
the conceptual definition of this operation from Remark~\ref{rem:advanced-HE} is
specified here as follows:
The circular rotation of $\cipher{v}$ by $i$ places yields $\rot_i(\cipher{v})=\cipher{v^\prime}$, where $v^\prime=[v_{j+i}]_{0\leq j < \batchsize-1}$.

Note that indices into the entries of vectors, the entries of matrices, or matrix diagonals are implicitly understood modulo $\batchsize$ here and in the following.
This implies, for example, that $s_{\batchsize-1,\batchsize+2} = s_{\batchsize-1, 2}$ and $S_{-i} = S_{\batchsize-i}$.

\section{Optimizations for homomorphic algorithms}\label{app:optimizations}

The basic algorithms for matrix-vector and matrix-matrix multiplication shown in \Cref{app:enc-matrix-vector-operations} can be improved upon in various ways.
In the following, we show one particular optimization that exploits structural features of matrix operands in matrix-vector multiplication relevant to our verification scheme.

In the special case that the function to be outsourced is linear (or affine, see \Cref{sec:lifting}), the lifted function is also linear and can be specified using a block-diagonal matrix.
When diagonal encoding is used, block-diagonal matrices exhibit sparse or, more specifically, banded structure.
In general, if $S\in\R^{\batchsize\times \batchsize}$ is banded with lower and upper bandwidth equal to $\beta\in[0, \frac{\batchsize-1}{2}]$,
\eqref{eq:matrix-vector-mult-diagonal} can be simplified by skipping the zero diagonals, resulting in:
\begin{align}
    \cipher{S v}
    &= \bigoplus_{i=-\beta}^{\beta} \cipher{S_i} \odot \rot_{i}(\cipher{v}).\label{eq:matrix-vector-mult-sparse}
\end{align}
This variant is more efficient than~\eqref{eq:matrix-vector-mult-diagonal} by a factor of $\frac{\batchsize}{2\beta+1}$ and can be applied to~\eqref{eq:matrix-matrix-mult-diagonal} as well.

%% file: manuscript.bbl
\begin{thebibliography}{22}
\setlength{\parskip}{0pt}
\setlength{\itemsep}{1pt plus 0.3ex}
\providecommand{\natexlab}[1]{#1}
\providecommand{\url}[1]{\texttt{#1}}
\providecommand{\urlprefix}{URL }
\expandafter\ifx\csname urlstyle\endcsname\relax
  \providecommand{\doi}[1]{doi:\discretionary{}{}{}#1}\else
  \providecommand{\doi}{doi:\discretionary{}{}{}\begingroup \urlstyle{rm}\Url}\fi

\bibitem[{Adamek et~al.(2024)Adamek, Binfet, Schlüter, and Schulze~Darup}]{adamek2024sysid}
Adamek, J., Binfet, P., Schlüter, N., and Schulze~Darup, M. (2024).
\newblock Encrypted system identification as-a-service via reliable encrypted matrix inversion.
\newblock In \emph{2024 IEEE 63rd Conf. Decis. Control (CDC)}, 4582--4588.

\bibitem[{Alexandru et~al.(2022)Alexandru, Burbano, Çeliktuğ, Gomez, Cardenas, Kantarcioglu, and Katz}]{alexandru2022anomaly}
Alexandru, A.B., Burbano, L., Çeliktuğ, M.F., Gomez, J., Cardenas, A.A., Kantarcioglu, M., and Katz, J. (2022).
\newblock Private anomaly detection in linear controllers: Garbled circuits vs. homomorphic encryption.
\newblock In \emph{2022 IEEE 61st Conf. Decis. Control (CDC)}, 7746--7753.

\bibitem[{Armknecht et~al.(2015)Armknecht, Boyd, Carr, Gjøsteen, Jäschke, Reuter, and Strand}]{armknecht2015guideFHE}
Armknecht, F., Boyd, C., Carr, C., Gjøsteen, K., Jäschke, A., Reuter, C.A., and Strand, M. (2015).
\newblock A guide to fully homomorphic encryption.
\newblock Cryptology {ePrint} Archive, Paper 2015/1192.

\bibitem[{Catalano and Fiore(2013)}]{catalano2013homomorphicMAC}
Catalano, D. and Fiore, D. (2013).
\newblock Practical homomorphic {MAC}s for arithmetic circuits.
\newblock In T.~Johansson and P.Q. Nguyen (eds.), \emph{Adv. Cryptol. -- EUROCRYPT 2013}, 336--352. Springer Berlin Heidelberg, Berlin, Heidelberg.

\bibitem[{Chatel et~al.(2024)Chatel, Knabenhans, Pyrgelis, Troncoso, and Hubaux}]{chatel2024veritas}
Chatel, S., Knabenhans, C., Pyrgelis, A., Troncoso, C., and Hubaux, J.P. (2024).
\newblock Veritas: Plaintext encoders for practical verifiable homomorphic encryption.
\newblock In \emph{Proc. 2024 ACM SIGSAC Conf. Comput. Commun. Secur.}, CCS '24, 2520–2534. Association for Computing Machinery, New York, NY, USA.

\bibitem[{Cheon et~al.(2017)Cheon, Kim, Kim, and Song}]{cheon2017ckks}
Cheon, J.H., Kim, A., Kim, M., and Song, Y. (2017).
\newblock Homomorphic encryption for arithmetic of approximate numbers.
\newblock In T.~Takagi and T.~Peyrin (eds.), \emph{Adv. Cryptol. -- ASIACRYPT 2017}, 409--437. Springer International Publishing, Cham.

\bibitem[{Dyer et~al.(2019)Dyer, Dyer, and Xu}]{dyer2019}
Dyer, J., Dyer, M., and Xu, J. (2019).
\newblock Practical homomorphic encryption over the integers for secure computation in the cloud.
\newblock \emph{Int. J. Inf. Secur.}, 18(5), 549–579.

\bibitem[{Fauser and Zhang(2020)}]{fauser2020}
Fauser, M. and Zhang, P. (2020).
\newblock Resilience of cyber-physical systems to covert attacks by exploiting an improved encryption scheme.
\newblock In \emph{2020 59th IEEE Conf. Decis. Control (CDC)}, 5489--5494.

\bibitem[{Fauser and Zhang(2021)}]{fauser2021}
Fauser, M. and Zhang, P. (2021).
\newblock Resilient homomorphic encryption scheme for cyber-physical systems.
\newblock In \emph{2021 60th IEEE Conf. Decis. Control (CDC)}, 5634--5639.

\bibitem[{Fauser and Zhang(2024)}]{fauser2024}
Fauser, M. and Zhang, P. (2024).
\newblock A secure resilient homomorphic encryption scheme for control systems.
\newblock \emph{IEEE Trans. Autom. Control}, 1--16.

\bibitem[{Gheitasi and Lucia(2020)}]{gheitasi2020finite}
Gheitasi, K. and Lucia, W. (2020).
\newblock A finite-time stealthy covert attack against cyber-physical systems.
\newblock In \emph{2020 7th Int. Conf. Control Decis. Inf. Technol. (CoDIT)}, volume~1, 347--352.

\bibitem[{Gheitasi and Lucia(2022)}]{gheitasi2022finiteCovertConstrained}
Gheitasi, K. and Lucia, W. (2022).
\newblock Undetectable finite-time covert attack on constrained cyber-physical systems.
\newblock \emph{IEEE Trans. Contr. Netw. Syst.}, 9(2), 1040--1048.

\bibitem[{Giraldo et~al.(2018)Giraldo, Urbina, Cardenas, Valente, Faisal, Ruths, Tippenhauer, Sandberg, and Candell}]{giraldo2018survey}
Giraldo, J., Urbina, D., Cardenas, A., Valente, J., Faisal, M., Ruths, J., Tippenhauer, N.O., Sandberg, H., and Candell, R. (2018).
\newblock A survey of physics-based attack detection in cyber-physical systems.
\newblock \emph{ACM Comput. Surv. (CSUR)}, 51(4), 1--36.

\bibitem[{Halevi and Shoup(2014)}]{halevi2014}
Halevi, S. and Shoup, V. (2014).
\newblock Algorithms in {HE}lib.
\newblock In J.A. Garay and R.~Gennaro (eds.), \emph{Adv. Cryptol. -- CRYPTO 2014}, 554--571. Springer Berlin Heidelberg, Berlin, Heidelberg.

\bibitem[{Johansson(2000)}]{johansson2000quadtank}
Johansson, K. (2000).
\newblock The quadruple-tank process: a multivariable laboratory process with an adjustable zero.
\newblock \emph{IEEE Trans. Control Syst. Technol.}, 8(3), 456--465.

\bibitem[{Katz and Lindell(2014)}]{Katz2014}
Katz, J. and Lindell, Y. (2014).
\newblock \emph{Introduction to modern cryptography, second edition}.
\newblock Chapman \& Hall/CRC Cryptography and Network Security Series. Chapman \& Hall/CRC, Philadelphia, PA, 2 edition.

\bibitem[{Marcolla et~al.(2022)Marcolla, Sucasas, Manzano, Bassoli, Fitzek, and Aaraj}]{marcolla2022surveyFHE}
Marcolla, C., Sucasas, V., Manzano, M., Bassoli, R., Fitzek, F.H.P., and Aaraj, N. (2022).
\newblock Survey on fully homomorphic encryption, theory, and applications.
\newblock \emph{Proc. IEEE}, 110(10), 1572--1609.

\bibitem[{Schl{\"u}ter et~al.(2023)Schl{\"u}ter, Binfet, and Schulze~Darup}]{schluter2023brief}
Schl{\"u}ter, N., Binfet, P., and Schulze~Darup, M. (2023).
\newblock A brief survey on encrypted control: From the first to the second generation and beyond.
\newblock \emph{Annu. Rev. Control}, 56, 100913.

\bibitem[{Schulze~Darup et~al.(2021)Schulze~Darup, Alexandru, Quevedo, and Pappas}]{Darup2021Survey}
Schulze~Darup, M., Alexandru, A.B., Quevedo, D.E., and Pappas, G.J. (2021).
\newblock Encrypted control for networked systems: An illustrative introduction and current challenges.
\newblock \emph{IEEE Control Syst. Mag.}, 41(3), 58--78.

\bibitem[{Smith(2011)}]{smith2011appropriating}
Smith, R.S. (2011).
\newblock A decoupled feedback structure for covertly appropriating networked control systems.
\newblock \emph{IFAC Proc. Vol.}, 44(1), 90--95.
\newblock 18th IFAC World Congress.

\bibitem[{Stabile et~al.(2024)Stabile, Lucia, Youssef, and Franzè}]{stabile2024verifyable}
Stabile, F., Lucia, W., Youssef, A., and Franzè, G. (2024).
\newblock A verifiable computing scheme for encrypted control systems.
\newblock \emph{IEEE Control Syst. Lett.}, 8, 1096--1101.

\bibitem[{Teixeira et~al.(2012)Teixeira, P\'{e}rez, Sandberg, and Johansson}]{teixeira2012attackModels}
Teixeira, A., P\'{e}rez, D., Sandberg, H., and Johansson, K.H. (2012).
\newblock Attack models and scenarios for networked control systems.
\newblock In \emph{Proc. 1st Int. Conf. High Confid. Netw. Syst.}, HiCoNS '12, 55–64. Association for Computing Machinery, New York, NY, USA.

\end{thebibliography}
